\documentclass{ws-ijseke}

\usepackage{url}
\usepackage[sort,compress]{cite}

\usepackage{graphicx}
\usepackage[table]{xcolor}
\usepackage{epstopdf} 
\usepackage{subfigure}
\usepackage{wrapfig}
\usepackage{bm}
\usepackage{float}

\usepackage{multirow}
\usepackage{tabularx}
\usepackage{makecell}
\usepackage{booktabs}

\usepackage{longtable}
\usepackage{arydshln}
\usepackage{mdframed}
\usepackage{verbatim}
\usepackage{lineno}

\definecolor{light-gray}{gray}{0.92}  

\newenvironment{gtheorem}%
  {\begin{mdframed}[backgroundcolor=light-gray]\begin{mdtheorem}{name}{label}}%
  {\end{mdtheorem}\end{mdframed}}

\newcommand{\tools}{\textit{Context-Encoded Representation}}
\newcommand{\tool}{\textit{Context-Encoded Change}}

\begin{document}

\markboth{Vu et. al.}
{Context-Encoded Code Change Representation for Automated Commit Message Generation}

%
\catchline{01}{01}{2003}{}{}
%

\title{Context-Encoded Code Change Representation for \\ Automated Commit Message Generation
}





\author{Thanh Trong Vu, Thanh-Dat Do, and Hieu Dinh Vo\footnote{Corresponding author}}

\address{Faculty of Information Technology, VNU University of Engineering and Technology, Hanoi, Vietnam\\
\email{\{19020626, 20020045, hieuvd\}@vnu.edu.vn}
}




\maketitle

\begin{abstract}
Changes in source code are an inevitable part of software development. They are the results of indispensable activities such as fixing bugs or improving functionality. Descriptions for code changes (commit messages) help people better understand the changes. However, due to a lack of motivation and time pressure, writing high-quality commit messages remains reluctantly considered. Several methods have been proposed with the aim of automated commit message generation.

However, the existing methods are still limited because they only utilise either the changed code or the changed code combined with surrounding statements. 

This paper proposes a method to represent code changes by combining the changed code and the unchanged code which have program dependence on the changed code. This method overcomes the limitations of current representations while improving the performance of 5/6 of state-of-the-art commit message generation methods by up to 15\% in METEOR, 14\% in ROUGE-L, and 10\% in BLEU-4.

\end{abstract}


\section{Introduction} 
Changes in source code are inevitable activities during the life cycle of software applications. Changes are made in order to fix bugs and improve functionality. Along with changes in source code, writing commit messages, which describe the changes, is an essential part of the software development process~\cite{girba2005developers,hassan2008road}.
Good commit messages help the code reviewers and maintainers easily understand the changes and the rationales behind them \cite{pascarella2018information, wessel2020expect,goodcommit-tian-2022, ma2023improving}. 
Although commit messages bring many benefits, in many cases, they are overlooked. This issue is affirmed by a survey conducted by Tian \textit{et. al.} ~\cite{goodcommit-tian-2022} on 1,600 randomly selected commit messages from five big open-source software projects. The results indicated that up to 44\% of the commit messages lacked crucial information regarding the nature of the code changes and the reasons for their implementation.  This means that the current commit messages contain insufficient necessary information to understand the code changes.  Also according to Tian \textit{et. al.}, a high-quality commit message must have two complete parts describing what has changed and why that change has occurred. 

Recently, several automated commit message generation methods have been proposed. 
%
%
Among these works, several studies rely on pre-defined rules or patterns~\cite{cortes-2014,shen-2016,huang2017mining}. Some studies apply information retrieval techniques to reuse commit messages for similar source code changes \cite{huang-2020, NNGen-liu-2018, CC2Vec,race2022}. Recently, Seq2Seq-based neural network models have been introduced to understand code changes and create high-quality commit messages \cite{NMT,CommitBERT-jung-2021, commitgen,CoDiSum, liu-2019,shia2022ecmg}. Although these approaches show promising results, they still have limitations which originate from the understanding and the representation of the code changes. 
Some of these methods represent code changes using only the changed code, which includes the added and removed statements. Because statements in changed code usually interact with the statements in unchanged code, it is unlikely to have high-quality commit messages if only changed code is used for message generation. Some other methods try to overcome this limitation by exploiting both the changed code and its surrounding statements. However, a statement positioned right before/after the changed code does not necessarily have a semantic relationship with the changed code. Using surrounding statements which have no semantic relationship with the changed code to generate the commit message should result in a low-quality one. To have high-quality commit messages, we need to consider the changed code and the part of the code that has a semantic relationship with it.

%

This paper proposes a context-based representation for code changes. The proposed representation aims at providing  information about changes in the program through statements that impact or are impacted by the changed code. Specifically, this representation is created by combining changed codes and unchanged codes that have program dependences on the changed code.
In addition, the proposed representation method also can be easily integrated with the current generation commit message methods. Our experiments on a dataset with 31,517 commits collected from 160 popular open source projects show that the proposed representation method can improve 5 out of 6 state-of-the-art methods for automated commit message generation. In particular, with the new change representation, the performance of these methods can be improved by up to 15\% in METEOR, 14\% in ROUGE-L, and 10\% in BLEU-4.

In brief, this paper makes the following contributions:
\begin{enumerate}
    \item \tools: A novel context-based code changes representation with substantial information about the changes.  
    \
    \item An extensive experimental evaluation showing the advantages of {\tools} over the state-of-the-art code change representations.
    \item A public dataset of 30K+  commits collected from 160 real-world projects, which can be used as a benchmark for evaluating related works.
\end{enumerate}

The rest of this paper is organized as follows. Section 2 provides the foundational knowledge about automated commit messages generation including an example illustrating our motivation and approach. Section 3 delineates the steps for the representation of code changes. Section 4 describes our evaluation methodology for the proposed code change representation. Section 5 presents and analyses the experiment results. We also discuss threats to validity of our work. The related works are in Section 6, before the conclusion.

\section{Background and Motivation}

This section presents the background knowledge of the automated commit message generation process. Subsequently, an exemplification of code changes accompanied by a corresponding commit message is provided. This example is used to illustrate our motivation and main ideals for representation of code changes.

\subsection{Automated Commit Message Generation}
\begin{figure}
    \centering
    \includegraphics[width=\columnwidth]{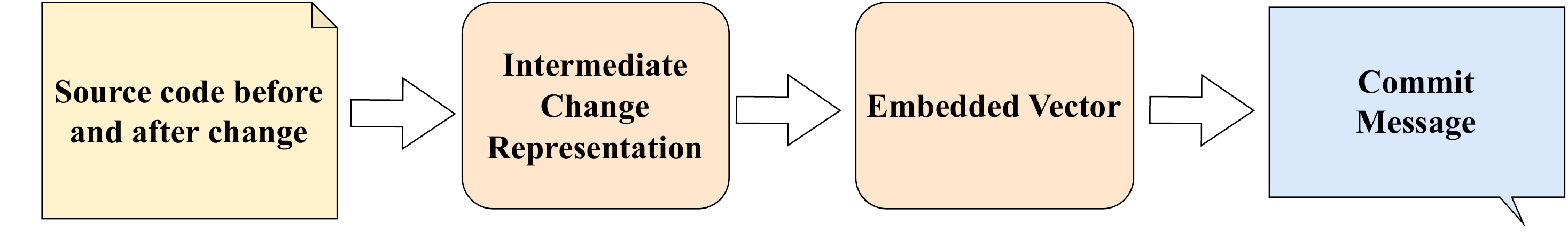}
    \caption{Typical process for automated commit message generation}
    \label{fig:pipeline_gen_mota}
\end{figure}

Figure \ref{fig:pipeline_gen_mota} presents the main steps in the process of automated commit message generation. First, for each change, two source code versions exist, including the versions before and after the change. Not all parts of source code is related to the change. Therefore, these two source code versions will be processed and transformed to create an intermediate change representation. 
At this step, existing researches may use either only changed code \cite{fira,CC2Vec,CommitBERT-jung-2021, CoDiSum} or both changed code and surrounding statements \cite{NNGen-liu-2018,NMT,ptr,race2022, commitgen,CommitBART}.

Subsequently, this intermediate change representation is transformed into an embedded vector representing the change's semantics. 
Various embedding approaches have been applied in existing researches.
Specifically, while Liu \textit{et. al.} utilized the ``Bags of words'' model \cite{bagsofwords} in NNGen \cite{NNGen-liu-2018}, recent studies have employed neural networks \cite{CC2Vec,ptr,CoDiSum,NMT,commitgen} and even pre-trained models \cite{CommitBERT-jung-2021,CommitBART}.



At the last step, commit messages are generated by using information retrieval techniques \cite{NNGen-liu-2018, CC2Vec} or, recently, neural networks \cite{ptr,CoDiSum,NMT,commitgen,CommitBERT-jung-2021,CommitBART}. Given a code change whose commit message needs to be generated, the approaches based on information retrieval techniques manage to find the most similar code change in the database. The generated commit message is the commit message of the most similar code change \cite{NNGen-liu-2018, CC2Vec}. Meanwhile, approaches using neural networks \cite{ptr,CoDiSum,NMT,commitgen,CommitBERT-jung-2021,CommitBART} take the embedded vectors as input for a machine translation process which returns the target commit message.

\begin{figure}
    \centering
    \includegraphics[width=0.8\columnwidth]{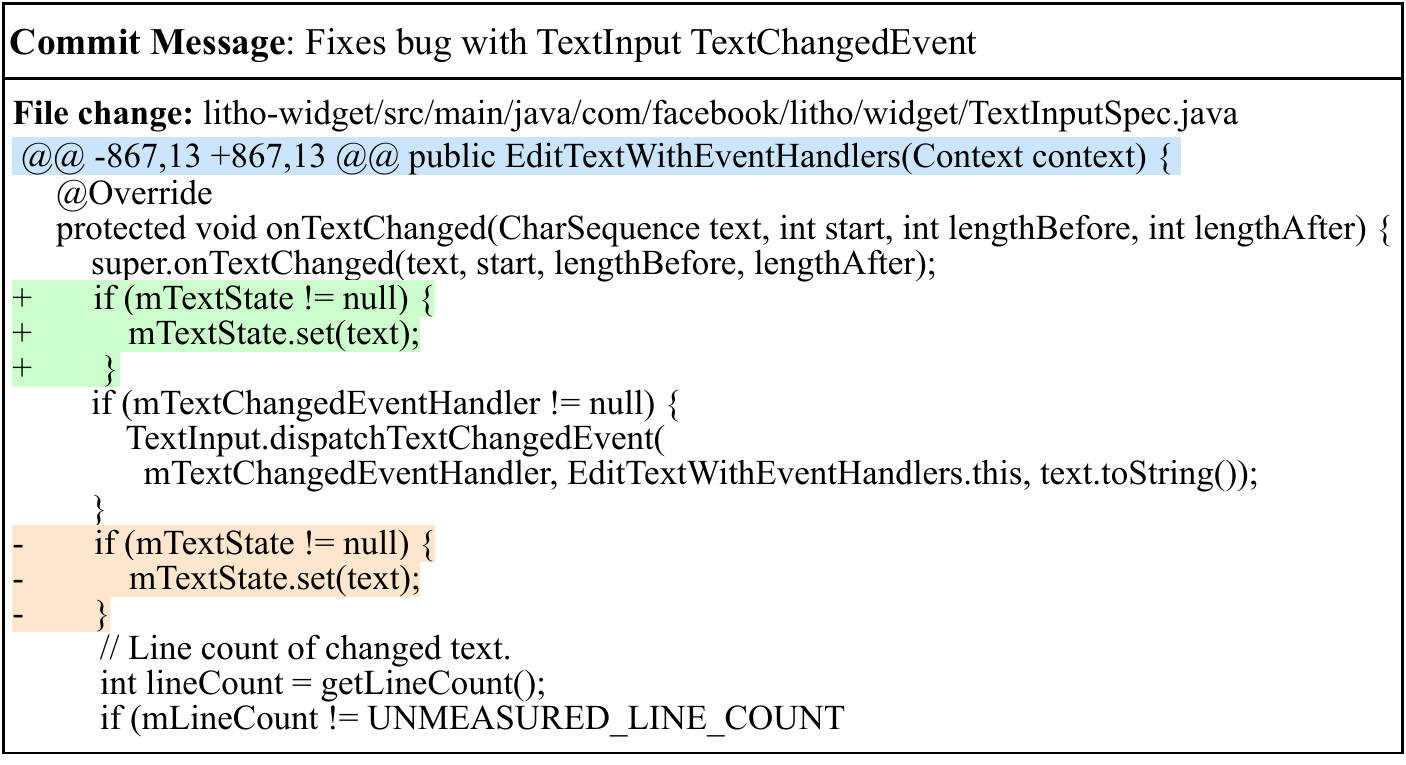}

    \caption{An example of a code change and its commit message.}
    \label{fig:example_mota}
\end{figure}
\subsection{Motivating Example}
Figure \ref{fig:example_mota} presents an example of a commit message for a simple change in the source code of the open-source project Litho \footnote{https://github.com/facebook/litho}. With this code change, if we only consider the changed code (as the existing studies \cite{CC2Vec, CoDiSum,CommitBERT-jung-2021,CommitBART,fira}), 
what we obtain is the information that the added code is identical to the removed code.
In fact, the changed statements may interact with unchanged parts of the program. Therefore, using only the changed code may not fully represent the change and may lead to difficulties in understanding the actual meaning of the change.
Moreover, the changed code may be similar between commits, but their purpose and meaning are entirely different because they are combined with different unchanged codes \cite{ctg}. Therefore, to accurately represent a code change, considering only the changed code seems not enough.

To address the issue of insufficient information about changes when relying solely on changed code, people may additionally use the statements surrounding the changed code \cite{NNGen-liu-2018,NMT,ptr,race2022, commitgen,CommitBART}.
Specifically, this approach takes a pre-defined number n of statements before and after the changed code.  In the example, if n is 3, the surrounding statements help to understand that the change moves the block \texttt{if (mTextState != null)} from behind to the front of the block \texttt{if (mTextChangedEventHandler != null)}. 
However, surrounding statements also may contain  undesirable ones. For example, the assignment \texttt{int lineCount = getLineCount() } and statement \texttt{ if (mLineCount != UNMEASURED\_LINE\_COUNT } have nothing to do with the changed code but still included as its surrounding.
These statements may significantly reduce the performance of commit message generation methods.

Therefore, to aim at generating high-quality commit messages, it is necessary to accurately understand the meaning of the change in the source code. This can be achieved by considering the changed code and the unchanged code which is the context of the changed code.

\section{Context-Encoded Code Change Representation}   
\begin{figure}
    \centering
    \includegraphics[width=\columnwidth]{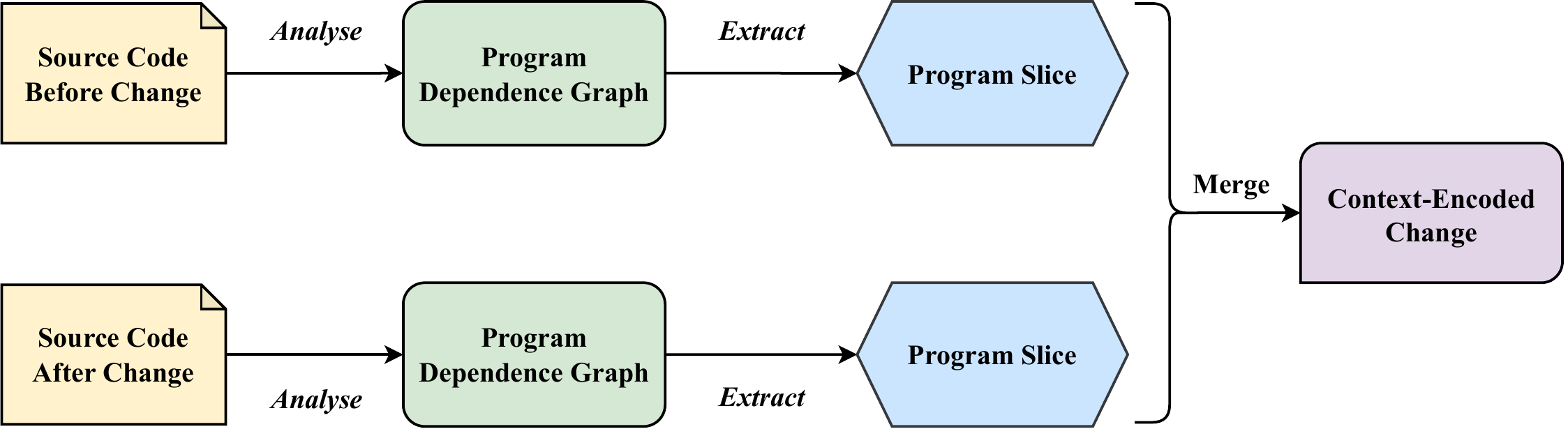}
    \caption{Main steps in constructing a context-encoded representation for a code change.}
    \label{fig:pipeline_create_malienket}
\end{figure}

%
In this section, we present the process of building code change representation named \tools, which combines the changed code and statements that are dependent on the changed code. Figure \ref{fig:pipeline_create_malienket} shows the main steps of our method including the construction of program dependence graph, program slice extraction, and context-encoded representation construction.

\subsection{Program Dependence Graph Construction}
\begin{figure}
    \centering
    \includegraphics[width=\columnwidth]{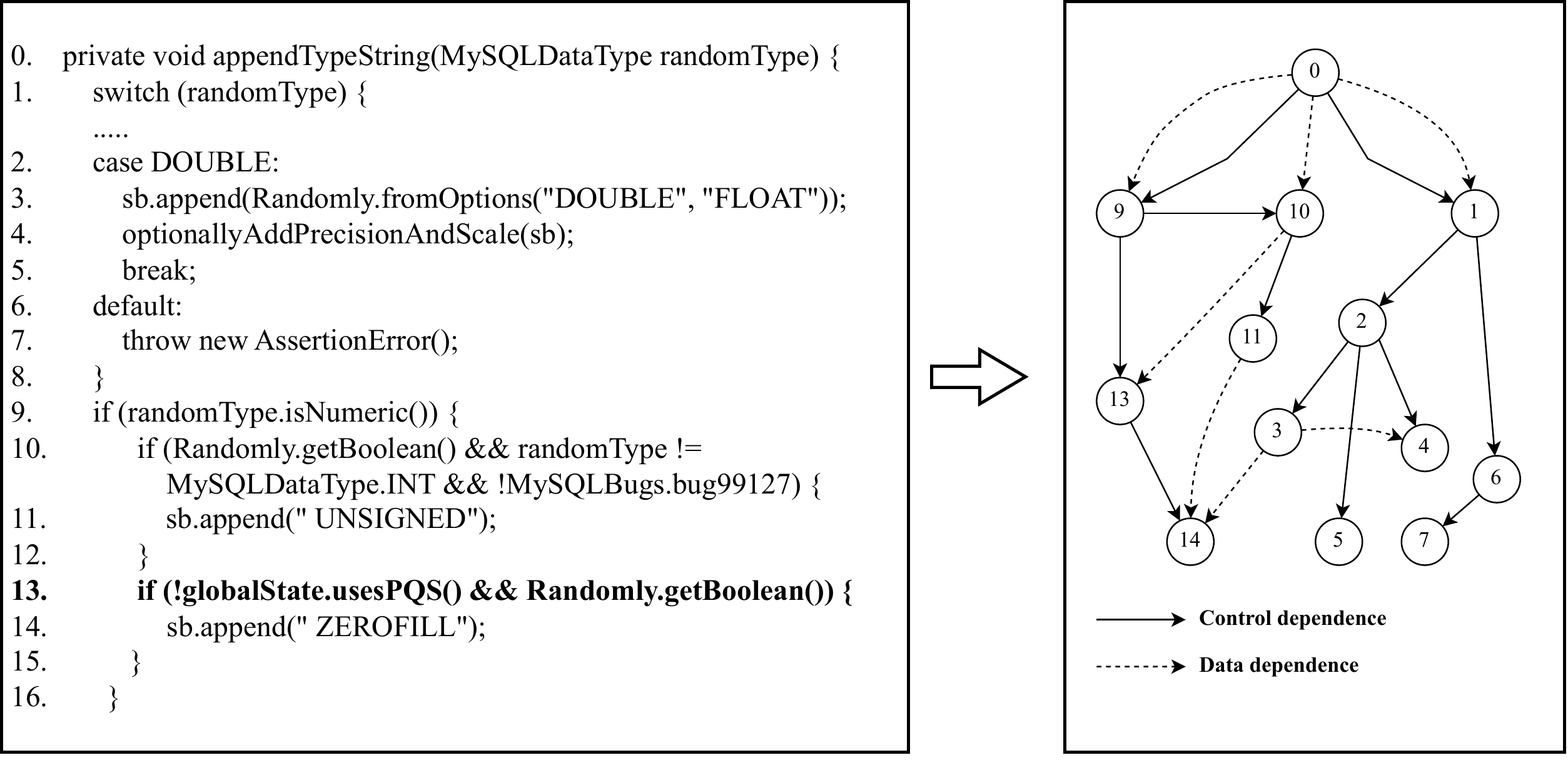}
    \caption{An example of a program dependence graph}
    \label{fig:example_program_ananesys}
\end{figure}

In this study,  to represent the dependencies between statements in the source code, we constructed a Program Dependence Graph (PDG). In a PDG, each node represents a statement while edges show relationships between statements.  As depicted in Figure \ref{fig:example_program_ananesys}, the graph effectively visualizes both data and control relationships between statements. Control dependence exists between two statements if one potentially prevents the execution of the other. Data dependence occurs when two statements declare, use, or reference the same variable.

\subsection{Program Slice Extraction}

From the program dependence graph, statements that have program dependences on the changed statements are extracted. 
Specifically, the proposed method extracts statements that have data dependences and control dependences on the changed code. 
In particular, we apply both backward and forward slicing. 
Furthermore, the inter-procedural slicing technique \cite{inter_slicing} is also applied
to ensure that statements outside the function will also be used to represent changes in the source code.

\subsection{ Context-Encoded Representation Construction}

\begin{figure}
    \centering
    \includegraphics[width=\columnwidth]{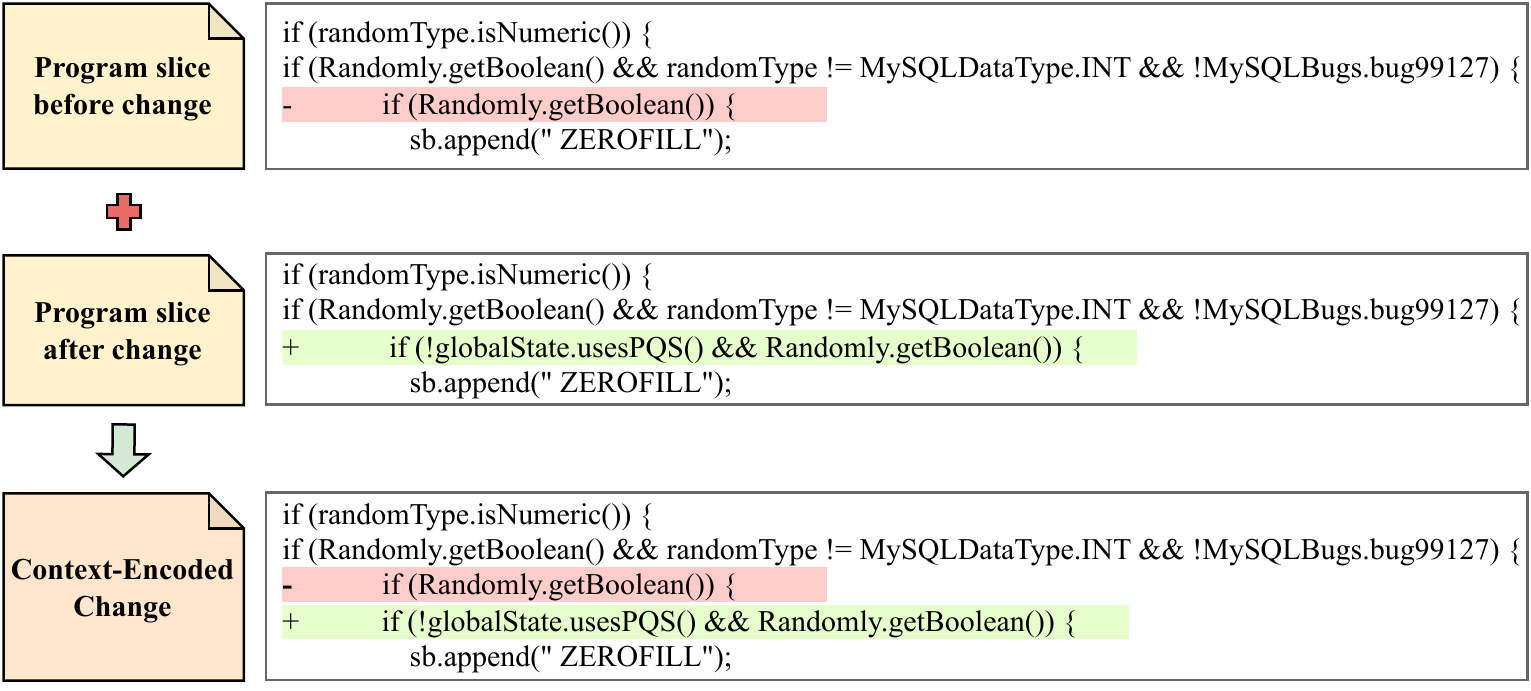}
    \caption{An example of building \tools}
    \label{fig:example_create_malienket}
\end{figure}

Figure \ref{fig:example_create_malienket} illustrates an example of combining program slices of the source code versions before and after the change to build the corresponding context-encoded representation of the change. Specifically, the context-encoded representation is created by combining added statements, removed statements, and unchanged codes that have program dependences on the changed code. In particular, unchanged statements that are the same between the two program slices will be merged. 
When combining the before and after versions of a change, the order of statements in the source code is preserved. Furthermore, changed statements are also marked to distinguish them from unchanged codes. Added statements are preceded by a \texttt{‘+’} character while the corresponding character for removed statements is \texttt{‘-’}. In particular, context-encoded representations are formatted as a sequence of statements that have program dependences on each other. This format is compatible with the representation of the existing methods for automated commit message generation (but different in content). 
This allows \tools\ easily be integrated with the existing methods.

\section{Evaluation Methodology}
To evaluate our approach, we seek to answer the following research questions:

\noindent\textbf{RQ1: Performance Improvement.} How does \tools\ improve the performance of the state-of-the-art methods for automated commit message generation \cite{CC2Vec,NNGen-liu-2018,CommitBERT-jung-2021,CommitBART,UniXcoder,CodeT5}?

\noindent\textbf{RQ2: Context Extraction Analysis.} How important are the dependences in \tools\ regarding its performance?

\noindent\textbf{RQ3: Dependency Depth Analysis.} How does the depth of dependences in \tools\ impact its performance?

\noindent\textbf{RQ4: Changes Complexity Analysis.} How does changed code’s complexity affect \tools? 

\subsection{Metrics}

The metrics used to evaluate the proposed method include BLEU-4, METEOR, and ROUGE-L, which are commonly used in machine translation researches\cite{CoDiSum,race2022, fira}. In addition, each metric represents a different aspect of evaluating text quality. BLEU-4 evaluates text quality up to 4 grams using uniform weights. Meanwhile, ROUGE-L evaluates text quality based on the longest common subsequence. Finally, METEOR evaluates the quality of generated text based on the harmonic mean of Precision and Recall of the unigram and their stemming and synonymy.

\subsection{Dataset}

The existing datasets \cite{NNGen-liu-2018} only provide information about the commit message along with the changed code and surrounding statements. This makes these datasets unsuitable for our evaluation. Therefore, we built a dataset by collecting all open-source Java projects which have at least 1,000 stars on Github (160 projects). Commits of these projects were then processed and filtered to meet the following criteria: having a sufficiently large change commit message length, being grammatically correct, and being well-evaluated by Yingchen’s model \cite{goodcommit-tian-2022}. Finally, 31,517 quality commits are retained (Table ~\ref{tab:commits}).

\begin{table}
\centering
\caption{Statistics of the dataset}\label{tab:commits}
\scriptsize
\begin{tabular}{lrrrr}\toprule
&\textbf{\#Commit} &\textbf{\#Changed} &\textbf{\#Changed/Commit} \\\midrule
\textbf{GraalVM} &1,763 &12,549 &7.12 \\
\textbf{Bazel} &1,493 &10,691 &7.16 \\
\textbf{Buck} &1,155 &8,410 &7.28 \\
\textbf{Loom} &1,109 &7,533 &6.79 \\
\textbf{Tomcat} &1,050 &7,295 &6.95 \\
\multicolumn{4}{c}{\textit{155 projects more}} \\\midrule
\textbf{Total} &31,517 &219,673 &6.97 \\
\bottomrule
\end{tabular}
\end{table}


Similar to previously published datasets \cite{CC2Vec, NMT, NNGen-liu-2018}, for each commit, 
we only consider the first sentence which usually summarises the content of the whole message. After that, the commit message is further cleansed by removing unique elements such as  Issue ID, commit ID, and URL.
In the next step, any commits with too many changed statements will be removed (more than 20 changes). Most of these changes are ``merger'' or ``rollback'' commits. In addition, commit messages that are too short (less than five words) are also removed because they are usually meaningless. Messages larger than 150 words are also dismissed. After that, the commit messages are checked for the `verb direct object' grammatical structure. Finally, they are fed into the deep learning model proposed by Yingchen and colleagues ~\cite{goodcommit-tian-2022} to classify whether they are good or not. Only good commit messages are retained.



\subsection{Experimental Setup} 
For data collection, we use Pydriller library\footnote{https://pydriller.readthedocs.io}. For analysing program dependencies, Joern~\cite{joern} is used. All experiments were performed on a server with an Intel Xeon (2) @ 2.00GHz CPU, 16 GB RAM, and an NVIDIA Tesla P100 PCIe 16GB GPU running Ubuntu 20.04.4 LTS x86\_64.

The maximum commit message size is 150 words, and the input size for generating commit messages is 512. Pre-trained models used in experiments are CodeT5, UniXcoder, and CodeBERT. Specifically for the pre-trained CodeT5 model, batch\_size, learning\_rate, and the number of epochs are 8, 5e-5, and 10, respectively. The pre-trained model initializing weights are \texttt{Salesforce/codet5-base}. For the pre-trained UniXcoder model, batch\_size, learning\_rate, and epoch are 12, 5e-5, and 10, respectively. The pre-trained model initializing weights are \texttt{microsoft/UniXcoder-base}. For other methods such as CommitBART \cite{CommitBART}, CommitBERT \cite{CommitBERT-jung-2021}, NNGen \cite{NNGen-liu-2018}, and CC2Vec \cite{CC2Vec}, the parameters are set with the default values.

\section{Experimental Results}
\label{sec:results}
\subsection{Performance Improvement (RQ1)}
\begin{table}
\centering
\caption{The impact of \tools\ on automated commit message generation}
\label{tab:performance}
\scriptsize
\begin{tabular}{llrrrr}\toprule
        \multicolumn{2}{c}{\textbf{Methods}} &\textbf{ METEOR} &\textbf{ROUGE-L} &\textbf{BLEU-4} \\\midrule

\multirow{2}{*}{\textbf{CommitBERT}} &Baseline &4.66 &10.81 &7.11 \\
&\tool &\textbf{5.36}  &\textbf{12.28} 
  & \textbf{7.83}   \\\midrule

\multirow{2}{*}{\textbf{CommitBART}} &Baseline &8.95 &15.28 &11.11 \\
&\tool &
\textbf{9.38} &\textbf{17.33} &\textbf{12.16} \\\midrule

\multirow{2}{*}{\textbf{CodeT5}} &Baseline &7.71 &15.17 &9.75 \\
&\tool &\textbf{8.46} &\textbf{16.29} &\textbf{10.54}  \\\midrule

\multirow{2}{*}{\textbf{UniXcoder}} &Baseline &5.45 &11.45 &7.64 \\
&\tool &\textbf{6.01} &\textbf{12.24} &\textbf{8.44} \\\midrule

\multirow{2}{*}{\textbf{NNGen}} &Baseline &4.04 &9.23 &5.83 \\
&\tool &\textbf{4.07} &\textbf{9.27} &\textbf{5.88} \\\midrule

\multirow{2}{*}{\textbf{CC2Vec}} &Baseline & 4.39 & 10.07 &6.21 \\
&\tool &4.31 &9.96 & 6.22 \\

\bottomrule
\end{tabular}
\end{table}
To evaluate the impact of \tools\ on the performance of state-of-the-art methods for automated commit message generation, we experiment with two scenarios: using the existing methods as they are and equipping the existing methods with \tools. The experiment involves different commit message generation methods, including methods based on information retrieval techniques such as NNGen \cite{NNGen-liu-2018} and CC2Vec \cite{CC2Vec} and methods based on neural network such as CommitBERT  \cite{CommitBERT-jung-2021} and CommitBART \cite{CommitBART}. We also investigate the impact of \tools\ on performance of pre-trained models such as CodeT5 \cite{CodeT5} and UniXcoder \cite{UniXcoder} in generating commit messages for code changes.

Table~\ref{tab:performance} describes the performance of automated commit message generation methods in two cases: the original version and the version with \tools. In general, \tools\ improves 5 out of 6 existing methods. The best performance improvement is with methods using pre-trained models. In particular, when applying \tools, for CommitBERT \cite{CommitBERT-jung-2021}, its performance is improved by up to 15\% in the METEOR metrics, 14\% in ROUGE-L, and 10\% in BLEU-4 metrics. Those figures for CommitBART \cite{CommitBART} are 5\%, 13\%, and 9\%, respectively. 
The performance of these methods is improved because when using \tools, pre-trained models can grasp the meaning of the change from program-dependent statements.

We can see that \tools\ does not improve the performance of CC2Vec \cite{CC2Vec} and NNGen \cite{NNGen-liu-2018}. These methods are based on information retrieval techniques. NNGen uses the “bags of words” model for embedding intermediate change representations.
Consequently, this method cannot represent the relationships between words and cannot utilize the program dependences provided by \tools. CC2Vec treats the added and removed source code separately before concatenating their embedded vectors into a single one representing changes in the source code. Therefore, when supplemented with unchanged code from \tools, this method cannot exploit the program dependences. In addition, due to the characteristics of information retrieval techniques, generated commit messages are actually taken from a pre-defined set of messages, the performance of these methods may reach their upper bounds.


\begin{gtheorem}
\textbf{Answer for RQ1}: \tools\ can improve 5/6 state-of-the-art methods for automated commit messages generation and improve performance by up to 15\%. 
\end{gtheorem}

\subsection{Context Analysis (RQ2)}
\begin{table}
\centering
\caption{Performance of \tools\ with different types of  dependence}\label{tab:impact_of_dependence}
\scriptsize
\begin{tabular}{lrrrr}\toprule
&\textbf{ METEOR} &\textbf{ROUGE-L} &\textbf{BLEU-4} \\\midrule
\textbf{Changed code} &7.71 &15.17 &9.75 \\
\textbf{Changed code + control dependence} &7.77 &15.16 &9.80 \\
\textbf{Changed code + data dependence} &8.25 &15.97 &10.41 \\
\textbf{Changed code + program dependence} &\textbf{8.46} &\textbf{16.29} &\textbf{10.54} \\
\bottomrule
\end{tabular}
\end{table}
In this experiment, we analyse the contribution of different code components in \tools\. Specifically, we generate commit messages with 04 different scenarios of \tools corresponding to 04 different ways of extracting context: (i) only the changed code, (ii) the changed code combined with its control dependence statements
(iii) the changed code combined with its data dependence statements, and (iv) the changed code combined with its program dependence (i.e. both control and data dependence) statements.

Table \ref{tab:impact_of_dependence} shows the contributions of different code components on \tools. We can see that \tools\ is at its best when changed code and program dependences are included. While data dependences are more helpful than control dependences in generating commit messages, we should use both of them (i.e. program dependences) to maximize the benefit from \tools. Specifically, compared with using only changed code, applying both changed code and program dependence helps enhance performance by 8\%, 7\%, and 10\% in BLEU-4, ROUGE-L, and METEOR, respectively.

\begin{gtheorem}
\textbf{Answer for RQ2}: The dependences within the source code has varying impacts on \tools. The method is at its best when both control dependences and data dependences are used.
\end{gtheorem}


\subsection{Dependence Depth Analysis (RQ3)}
\begin{table}[!htp]\centering
\caption{Performance of \tools\ with different context extraction depths}\label{tab:impact_of_deep}
\scriptsize
\begin{tabular}{lrrrr}\toprule
\textbf{\#Deep level} &\textbf{ METEOR} &\textbf{ROUGE-L} &\textbf{BLEU-4} \\\cmidrule{1-4}
\textbf{Deep 1} &8.31 &15.97 &10.42 \\
\textbf{Deep 2} &8.30 &16.00 &10.35 \\
\textbf{Deep 3} &8.46 &16.29 &10.54 \\
\textbf{Deep 4} &8.26 &15.95 &10.36 \\
\textbf{Deep 5} &7.68 &15.08 &9.76 \\
\bottomrule
\end{tabular}
\end{table}
To analyse the impact of the depth of dependences on \tools\, we equip it with different levels of dependence depth. For example, in Figure \ref{fig:example_program_ananesys}, when the depth is 1, the statements that have program dependences with statement 13 are 9, 10, and 14. Meanwhile, when the depth is 2, the statements dependent on the statement 13 are 0, 9, 10, and 14. 
In our experiment, the depth varies from one to five.

Table \ref{tab:impact_of_deep} describes the impact of dependence depths on generating commit messages. 
The results show that by increasing the depth of the dependences \tools\ may provide better performance for generating commit messages. However, \tools\ is at its best when dependence depth is 3. Above this value, increasing the dependence depth will result negative impact on commit message generation. 

\begin{gtheorem}
\textbf{Answer for RQ3}: The depth of program dependences slightly impacts the performance of \tools. 
\end{gtheorem}

\subsection{Changes Complexity Analysis (RQ4)}
\begin{table}
\centering
\caption{Performance of \tools\ with different input complexity}\label{tab:impact_of_change_size}
\scriptsize
\begin{tabular}{lrrrrr}\toprule
\textbf{\#Number of changed statements} &\textbf{ METEOR} &\textbf{ROUGE-L} &\textbf{BLEU-4} \\\midrule
\textbf{From 1 to 5}  &8.35 &16.05 &10.61 \\
\textbf{From 5 to 10}   &8.33 &16,00 &10.18 \\
\textbf{From 10 to 15}   &8.20 &15.76 &10.27 \\
\textbf{Over 15}   &8.03 &15.14 &9.89 \\
\bottomrule
\end{tabular}
\end{table}
We also evaluate the impact of complexity in each change on the quality of the corresponding generated commit message. Specifically, we evaluate the quality of generated commit messages for changes including 1 to 5 changed statements, 5 to 10 changed statements, 10 to 15 changed statements, and more than 15 changed statements.

Table \ref{tab:impact_of_change_size} shows the impact of different input sizes on \tools. \tools\ reaches its best performance when the number of changed statements is between 1 and 5. Its performance then gradually decreases as the number of changed statements increases. The reason is that when the number of changes increases, it is more difficult to grasp the meaning of them fully. 
\begin{gtheorem}
\textbf{Answer for RQ4}: The complexity of the change has a significant impact on the performance of \tools. The more complex the change, the lower the performance of \tools.

\end{gtheorem}

\subsection{Time Complexity}
In this paper, we use Joern to analyse the program, it takes about 1 second to explore a commit and 0.1 seconds to build the \tool. Compared to other representation techniques, there is not much difference in the training and evaluation process when using \tools. With the commit message generation methods using pre-trained models such as CodeT5 \cite{CodeT5} or CommitBERT \cite{CommitBERT-jung-2021}, the training time when using \tools\  and the default method both take around 500 minutes and 540 minutes, respectively.

\subsection{Threats to Validity}
The main threats to the validity of our work consist of internal, construct, and external threats.

\textbf{Threats to internal validity} include the influence of the method used to extract program dependencies. To reduce this threat, we use Joern~\cite{joern} code analyzer, which is a widely-used code analyzer. Another threat mainly lies in the correctness of the implementation of our approach. To reduce such threats, we carefully reviewed our code and made it public~\cite{c3} so that other researchers can double-check and reproduce our experiments.

\textbf{Threats to construct validity} relate to the suitability of our evaluation procedure. We used \textit{BLEU-4}, \textit{METEOR}, and \textit{ROUGE-L}. They are the widely-used evaluation measures for evaluating commit messages generated by the existing techniques~\cite{commitgen, NMT, ptr, fira, CoDiSum, CommitBERT-jung-2021}. 
Besides, a threat may come from the adaptation of the existing commit message generation approaches. To mitigate this threat, we directly obtain the original source code from the GitHub repositories of the studied techniques. Also, we use the same hyperparameters as in the original papers~\cite{CC2Vec,NNGen-liu-2018,CommitBERT-jung-2021,CommitBART,UniXcoder,CodeT5}. 

\textbf{Threats to external validity} mainly lie in the construction of our dataset. 
%
%
%
To reduce the impact of this threat, we applied the same data collection procedure as in the existing commit message generation approaches~\cite{commitgen, NMT, ptr, fira, CoDiSum, CommitBERT-jung-2021} and collect the commits in high-quality Java projects. 
Moreover, our experiments are conducted on only the commits in Java projects. Thus, the results could not be generalized for other programming languages. In our future work, we plan to conduct more experiments to validate our results to other languages.

\section{Related Work}

\textbf{Code Change Representation}. Our approach is related to code change representation approaches. Before using in a specific task, code changes could be transformed into several forms such as sequence-based~\cite{sequence-based}, tree-based~\cite{structure-based}, and graph-based~\cite{cpathminer,ctg}. Additionally, several techniques have been proposed to learn how to represent changes~\cite{slice-based, CC2Vec, fira,commit2vec}.
These studies differ from our work as our approach considers the changed code sequences in relation to the related unchanged code and is designed to directly integrate into the existing commit message generation methods.

\textbf{Commit message generation}. Our work is also related to learning-based commit message generation methods. Recently, a number of neural network-based approaches~\cite{commitgen, NMT, ptr, fira, CoDiSum, CommitBERT-jung-2021} have been used to understand the semantics of code changes in the commit and translate them into commit messages. NMTGen~\cite{commitgen} and CommitGen \cite{NMT} treat code changes as pure text and use Seq2Seq neural networks with different attention mechanisms to translate them into commit messages. CoDiSum \cite{CoDiSum} extracts the structure and semantics of code and creates a multi-layer bidirectional GRU model to better understand the representation of code changes. CommitBERT \cite{CommitBERT-jung-2021} uses CodeBERT \cite{CodeBERT}, a pre-trained language model for code, to understand the semantics of code changes and applies a transformer-based decoder \cite{transformer-vaswani-2017} to generate commit messages. 
%
Our study differs from these studies as instead of focusing on end-to-end commit message generation, our method is designed to integrate into these existing commit message generation techniques. Additionally, as shown in Section~\ref{sec:results}, our approach and those approaches can be applied together to generate better commit messages.

\textbf{Learning-based SE approaches}.
Several \textbf{learning-based approaches} have been proposed for specific SE tasks including code recommendation/suggestion~\cite{icse20, naturalness, allamanis2016convolutional}, program synthesis~\cite{amodio2017neural,gvero2015synthesizing},  static analysis warnings ~\cite{ngo2021ranking, vu2022using}, pull request description generation~\cite{hu2018deep,liu2019automatic}, code summarization~\cite{iyer2016summarizing,mastropaolo2021studying,wan2018improving}, code clones~\cite{li2017cclearner}, fuzz testing\cite{godefroid2017learn}, bug detection~\cite{oppsla19}, and program repair~\cite{jiang2021cure,ding2020patching}.


\section{Conclusion}
Commit message is very important in the field of software development. Numerous methods for automated commit message generation have been proposed, yielding remarkable outcomes. Nonetheless, the representation of code changes in these methods does not fully cover information about changes and is often accompanied by noise.
This paper presents a new technique for representing code changes by exploiting unchanged code that had program dependence on the changed code. Experiment results on a dataset collected from 160 open-source projects on Github, including 31,517 commits, show that by using the proposed representation method, the performance of 5 out of 6 state-of-the-art automated commit message generation methods can be improved. In particular, the performance of these methods can be improved by up to 15\% METEOR compared to using current representations for code changes.


\bibliographystyle{ws-ijseke}
\bibliography{sample}

\end{document}